\documentstyle[12pt,epsfig]{article}

\newcommand{\beq}{\begin{equation}}
\newcommand{\eeq}{\end{equation}}
\newcommand{\beqa}{\begin{eqnarray}}
\newcommand{\eeqa}{\end{eqnarray}}
\newcommand{\al}{\alpha}
\newcommand{\as}{\alpha_s}
\newcommand{\api}{\frac{\alpha_s}{\pi}}
\newcommand{\ba}{\begin{array}}
\newcommand{\ea}{\end{array}}
\newcommand{\msbar}{\overline{\mbox{MS}}}

\newcommand{\D}{\displaystyle}

\newcommand{\ep}{\epsilon}
\newcommand{\drho}{\delta\rho}
\newcommand{\mbar}{\bar{m}}
\newcommand{\lbar}{l}
\newcommand{\los}{\log\frac{\mu^2}{M_t^2}}
\newcommand{\sineff}{\sin^2\!\Theta_{\mbox{\footnotesize eff}}}

\begin{document}

\begin{titlepage}
\noindent
%
%
\hfill TTP95--03\\
\mbox{}
\hfill hep-ph/9502291\\
\mbox{}
\hfill February 1995   \\

\vspace{1cm}

\begin{center}
\begin{LARGE}
 {\bf Corrections of order ${\cal O}(G_F M_t^2 \as^2)$ to the \\
  $\rho$ parameter}
\end{LARGE}

\vspace{.8cm}

\begin{large}
 K.G.~Chetyrkin$^{a,b}$,
 J.H.~K\"uhn$^{b}$,
 M.~Steinhauser$^{b}$
\end{large}

\vspace{1.5cm}
\begin{itemize}
\item[$^a$]
   Institute for Nuclear Research\\
   Russian Academy of Sciences, 60th October Anniversary Prospect 7a,
   Moscow, 117312, Russia
\item[$^b$]
   Institut f\"ur Theoretische Teilchenphysik\\
   Universit\"at Karlsruhe, Kaiserstr. 12,    Postfach 6980,
   D-76128 Karlsruhe, Germany\\
\end{itemize}

\vspace{2cm}

\begin{abstract}
\noindent
The three-loop QCD corrections to the $\rho$ parameter from top and
bottom quark loops are calculated.
The result differs from the one recently calculated
by Avdeev et al. As function of the pole mass the numerical value is given by
$\drho=\frac{3G_F M_t^2}{8\sqrt{2}\pi^2}(1- 2.8599\, \api
- 14.594\, (\api)^2 )$.
\end{abstract}


\vfill
\end{center}
\end{titlepage}

\renewcommand{\arraystretch}{2}

\section{Introduction}

The precision of electroweak observables measured at LEP, SLC
and the TEVATRON has stimulated a variety of theoretical calculations.
These are required to match the experimental accuracy and to pin down the
parameters of the Standard Model, in particular the top and the Higgs
mass and to search for new physics.

A cornerstone in this analysis is the evaluation of top quark contributions
to the $\rho$ parameter. With the high value of $M_t$ as suggested
by the CDF-collaboration \cite{CDF} and the strong sensitivity
of $\rho$ to small variations of $M_t$
through the quadratic dependence
\cite{Vel77}
precise theoretical predictions become mandatory.

Top mass corrections to the $Zb\bar{b}$ vertex, the only other place
with a strong  dependence on $M_t$
\cite{zbbew,zbbqcd},
are specific to this reaction, while those to the $\rho$ parameter
enter numerous relations between observables. This is the second
justification for a precision calculation.

In addition to the ${\cal O}(G_F M_t^2 \as)$ two loop contribution
\cite{DjoVer87Djo88Kni90}
(for a related calculation based on dispersion relations see
\cite{KniKueStu88},
for a recent review see
\cite{Kni94}),
the two-loop electroweak corrections have also been evaluated, for
vanishing \cite{VdBHoo97}
and even for arbitrary $M_H$
\cite{Bar92FleTarJeg93}.

In this paper the three-loop result of order $G_F M_t^2 \as^2$ is
presented. A similar calculation has been performed by
Avdeev, Fleischer, Mikhailov and Tarasov
\cite{AvdFleMikTar94}. However, our calculation disagrees with their
formula. In the following section 2 details of the calculation will be
described, in section 3 the result will be presented and a brief numerical
discussion given.

\section{The calculation}

Quantum corrections to the $\rho$ parameter can be connected to the gauge
boson self-energies through
\beq
\drho = \frac{\Pi^Z(0)}{M_Z^2} - \frac{\Pi^W(0)}{M_W^2}.
\eeq
Here $\Pi(0)$ denotes the transverse part of the polarisation tensor
$\Pi^{\mu\nu}(q)$ for vanishing momentum $q$.
The evaluation of these self-energy diagrams
is performed for $m_b=0$ and within the
framework of dimensional regularisation.
Large intermediate expressions are treated with the help of FORM 2.0
\cite{Ver91}.
For $D\not=4$
anticommuting $\gamma_5$ was used, except for the double triangle diagram.
In order to evaluate this diagram, which is related to the axial anomaly,
the definition of 't Hooft and Veltman
\cite{tHoVel72},
formalized in
\cite{BreMai77}, was applied.
Its contribution is finite and the result coincides with a previous calculation
\cite{AnsDomLea93}
where $D=4$ from the very beginning.
A covariant gauge with arbitrary gauge parameter
for the gluon propagator was chosen.

The tadpole integrals required to calculate the one- or two-loop
corrections are easy to evaluate even for arbitrary powers
of the propagators.
This does not hold true for the three-loop case.
After performing the traces the reduction to scalar integrals is performed by
decomposing the scalar products of the numerator in appropriate combinations
of the denominator. Subsequently,
recurrence relations provided by the
integration-by-parts (IP) method
\cite{CheTka81}
are used in order to reduce every
scalar Feynman integral to a small number of so-called
master diagrams which have to be calculated explicitly.
The IP method was first applied in
\cite{Bro92} to three-loop tadpole integrals. There the subclass of
those diagrams which contain a continuous massive quark line and which
are relevant for the $Z$ boson self-energy was considered.
For the calculation of the
$\rho$ parameter the method has to be extended to a second class of
integrals originating from the $W$ self-energies. One thus arrives
at three master integrals
\begin{eqnarray*}
\ba{lcl}
\D \int\!\!\!\int\!\!\!\int\frac{d^D\! p\,d^D\! k\,d^D\! l}{(\pi^{D/2})^3}
   \frac{1}{l^2 k^2 (M^2+p^2)(M^2+(p+k)^2)(M^2+(p+l)^2)(M^2+(p+k+l)^2)}
&&
\\ \D =\left(\frac{\mu^2}{M^2}\right)^{3\ep}\left[
        \frac{2}{\ep}\zeta(3) +  6\zeta(3) - 9\zeta(4) + 2 B_4
                       \right]
\\
\D \int\!\!\!\int\!\!\!\int\frac{d^D\! p\,d^D\! k\,d^D\! l}{(\pi^{D/2})^3}
             \frac{1}{(M^2+p^2)(M^2+k^2)(M^2+l^2)(p+k+l)^2}
&&
\\
\D =\left(\frac{\mu^2}{M^2}\right)^{3\ep} M^4 \left[
         \frac{1}{\ep^3}
       + \frac{15}{4\ep^2}
       + \frac{3}{2\ep}\zeta(2) + \frac{65}{8\ep}
       + \frac{81}{4} S_2 - \zeta(3) + \frac{45}{8}\zeta(2) + \frac{135}{16}
          \right]
\\
\D \int\!\!\!\int\!\!\!\int\frac{d^D\! p\,d^D\! k\,d^D\! l}{(\pi^{D/2})^3}
   \frac{1}{l^2 k^2 p^2(M^2+(p+k)^2)(M^2+(p+l)^2)(M^2+(p+k+l)^2)}
&&
\\
\D =\left(\frac{\mu^2}{M^2}\right)^{3\ep} \left[
                  \frac{2}{\ep}\zeta(3) + D_3
                                         \right]
\ea
\end{eqnarray*}
(The same variables and conventions as those of
\cite{AvdFleMikTar94} are adopted. Following standard $\msbar$ practice
we discarded terms proportional to $\gamma_E$ and $\log 4\pi$
in the r.h.s.)
The first master integral has been calculated analytically in
\cite{Bro92}; the results for the last two integrals are given
\cite{AvdFleMikTar94} ($D_3$ is presently only known numerically.).
We checked the result for $B_4$ numerically. For $S_2$ we reproduced
the analytical result. The evaluation of $D_3$ is described in
\cite{FleTar94}.
Employing a different method, we
obtain the result given below, which is consistent with \cite{FleTar94}.
The values for the
constants $B_4$, $S_2$ and $D_3$ are as follows:
\begin{eqnarray*}
B_4 &=& 16\, \mbox{Li}_4\left(\frac{1}{2}\right)
      + \frac{2}{3}\log^4 2
      - \frac{2}{3}\pi^2\log^2 2
      - \frac{13}{180}\pi^4
     = -1.76280\ldots \\
S_2 &=& \frac{4}{9\sqrt{3}} \mbox{Cl}_2\left(\frac{\pi}{3}\right)
     = 0.260434\ldots\\
D_3 &=& -3.02700\ldots
\end{eqnarray*}

\section{Results and Discussion}

After performing mass and charge renormalization in the
$\msbar$ scheme the following result for the $W$ boson
propagator is obtained:
\beqa
\lefteqn{\Pi^W(0) = 12 x_t M_W^2}                  \label{piw}
\\ & &
         \Bigg\{ - \frac{1}{2\ep} -\frac{1}{4} - \frac{1}{2} \lbar
       + \frac{\as}{4\pi} C_F \bigg(\frac{3}{2\ep^2} - \frac{5}{4\ep}
              - \frac{13}{8} + \zeta(2) -  \lbar - \frac{3}{2} \lbar^2 \bigg)
\nonumber\\ & &
       + \left(\frac{\as}{4\pi}\right)^2\Bigg[
         C_F  \bigg(   - \frac{44}{3} - \frac{243}{2} S_2
              + \frac{92}{3} \zeta(3) + \frac{7}{3}\zeta(2)\bigg)
\nonumber\\ & &
       + C_F^2 \bigg(  - \frac{3}{\ep^3} + \frac{3}{\ep^2}
          - \frac{6}{\ep} \zeta(3)
          + \frac{119}{24\ep} + \frac{1025}{72} + \frac{1053}{4} S_2 - D_3
          - 18 \zeta(3) \lbar - \frac{379}{3} \zeta(3)
\nonumber\\ & &
\phantom{C_F C_A}
          + 26 \zeta(4) + 6 \zeta(2) \lbar
          + \frac{259}{18} \zeta(2) + \frac{95}{8} \lbar
          + \frac{21}{4} \lbar^2
          - 3 \lbar^3 - 8 B_4 \bigg)
\nonumber\\ & &
         +C_F C_A \bigg(-\frac{11}{6\ep^3} + \frac{83}{12\ep^2}
         + \frac{3}{\ep}\zeta(3) -
           \frac{77}{12\ep} - \frac{869}{48} - \frac{1053}{8} S_2
         + \frac{1}{2} D_3
         + 9 \zeta(3) \lbar
\nonumber\\ & &
\phantom{C_F C_A}
         + 47 \zeta(3)  - 21 \zeta(4) + \frac{11}{3} \zeta(2) \lbar
         + \frac{73}{6} \zeta(2)
         - \frac{137}{8} \lbar
         - \frac{119}{12} \lbar^2 - \frac{11}{6} \lbar^3
         + 4 B_4 \bigg)
\nonumber\\ & &
       + C_F n_f \bigg( \frac{1}{3\ep^3} - \frac{5}{6\ep^2} + \frac{2}{3\ep}
       + \frac{73}{24} + 4 \zeta(3)
       - \frac{2}{3} \zeta(2) \lbar
       - \frac{7}{3} \zeta(2)
       + \frac{9}{4} \lbar + \frac{7}{6} \lbar^2 + \frac{1}{3} \lbar^3 \bigg)
\nonumber
       \Bigg]
       \Bigg\}
\eeqa
The coefficient of $C_F^2\zeta(4)$ differs from the recent result
of \cite{AvdFleMikTar94}. Its value $26$ has to be compared with $88/5$.
This leads to a significant modification of the numerical predictions
to be discussed below. In this expression $n_f=6$ denotes the total number
of quark species and $\lbar\equiv\log\mu^2/\mbar_t^2$. The result is expressed
in terms of the $\msbar$ renormalized top mass $\mbar_t(\mu^2)$.
The variable $x_t$ is defined as
\beq
x_t(\mu^2) = \frac{G_F \mbar_t^2(\mu^2)}{8\sqrt{2}\pi^2}.
\eeq
{}From now on the explicit $\mu$-dependence both in $x_t$ and
$\alpha_s$ will be suppressed. From the context it should be evident
which scale is adopted.

For the $Z$ boson propagator the following result is obtained
\beqa
\lefteqn{\Pi^Z(0) = 12x_t M_Z^2}                           \label{piz}
\\ & &
         \Bigg\{
       - \frac{1}{2\ep} - \frac{1}{2} \lbar
       + \frac{\as}{4\pi}
         C_F \bigg(   \frac{3}{2\ep^2} - \frac{5}{4\ep}
                     - \frac{1}{8} + \frac{1}{2} \lbar - \frac{3}{2} \lbar^2
             \bigg)
\nonumber\\ & &
       + \left(\frac{\as}{4\pi}\right)^2\Bigg[
         C_F \bigg(  - 2 - 12 \zeta(3) \bigg)
\nonumber\\ & &
       + C_F^2  \bigg( - \frac{3}{\ep^3} + \frac{3}{\ep^2}
                 - \frac{6}{\ep} \zeta(3) + \frac{119}{24\ep}
                 + \frac{51}{16} - 18 \zeta(3) \lbar
                 - 36 \zeta(3) + 27 \zeta(4) + \frac{101}{8} \lbar
\nonumber\\ & &
\phantom{C_F C_A}
                  + \frac{39}{4} \lbar^2 - 3 \lbar^3
                 - 6 B_4 \bigg)
\nonumber\\ & &
         C_F C_A \bigg( - \frac{11}{6\ep^3} + \frac{83}{12\ep^2}
                      + \frac{3}{\ep} \zeta(3) - \frac{77}{12\ep}
                      + 3 + 9 \zeta(3) \lbar
                      + \frac{28}{3} \zeta(3) - \frac{27}{2} \zeta(4)
                      - \frac{85}{24} \lbar
\nonumber\\ & &
\phantom{C_F C_A}
                       - \frac{43}{6} \lbar^2 - \frac{11}{6} \lbar^3
                      + 3 B_4 \bigg)
\nonumber\\ & &
       + C_F n_f \bigg(  \frac{1}{3\ep^3} - \frac{5}{6\ep^2}
                       + \frac{2}{3\ep} - \frac{1}{12} + \frac{8}{3} \zeta(3)
                       + \frac{5}{12} \lbar
                       + \frac{2}{3} \lbar^2 + \frac{1}{3} \lbar^3 )
\nonumber
        \Bigg]
        \Bigg\}.
\eeqa
in agreement with \cite{AvdFleMikTar94}.
Both $\Pi^W$ and $\Pi^Z$ are independent of the gauge parameter.
As expected this holds true even before mass and charge renormalization are
performed.

Eqs. (\ref{piw}) and (\ref{piz}) immediately lead to $\drho$ in terms
of the $\msbar$ renormalized top mass. The poles which are still
present in the $Z$ and $W$ self-energies individually cancel.
\beqa
\lefteqn{\drho_{\scriptsize\msbar} = 3 x_t}        \label{drhoms}
\\ &&
        \Bigg\{ 1
            + \frac{\as}{4\pi} C_F \bigg( 6 - 4 \zeta(2) + 6\lbar \bigg)
\nonumber \\ & &
       + \left(\frac{\as}{4\pi}\right)^2\Bigg[
         C_F  \bigg( \frac{152}{3} + 486 S_2 - \frac{512}{3} \zeta(3)
                     - \frac{28}{3} \zeta(2) \bigg)
\nonumber \\ & &
       + C_F^2 \bigg(  - \frac{1591}{36} - 1053 S_2 + 4 D_3
                       + \frac{1084}{3} \zeta(3)
                       + 4 \zeta(4) - 24 \zeta(2) \lbar
\nonumber\\ & &
\phantom{C_F C_A}
                       - \frac{518}{9} \zeta(2) + 3 \lbar
                       + 18 \lbar^2 + 8 B_4 \bigg)
\nonumber \\ & &
         C_F C_A \bigg( \frac{1013}{12} + \frac{1053}{2} S_2 - 2 D_3
                      - \frac{452}{3} \zeta(3) + 30 \zeta(4)
                      - \frac{44}{3} \zeta(2) \lbar - \frac{146}{3} \zeta(2)
\nonumber\\ & &
\phantom{C_F C_A}
                      + \frac{163}{3} \lbar
                      + 11 \lbar^2 - 4 B_4 \bigg)
\nonumber \\ & &
       + C_F n_f  \bigg(  - \frac{25}{2} - \frac{16}{3} \zeta(3)
                          + \frac{8}{3} \zeta(2) \lbar
                          + \frac{28}{3} \zeta(2)
                          - \frac{22}{3} \lbar
                          - 2 \lbar^2 \bigg)
             \Bigg]
             \Bigg\}
\nonumber
\eeqa
With the help of the relation between the OS and running top mass
\cite{BroGraGraSch90} for $\mu^2=M_t^2$
\beqa
  \mbar_t(M_t^2) &=& M_t \Bigg[ 1 - \api  C_F
                     + \left(\api\right)^2 \Bigg( C_F n_f \bigg(
                       \frac{1}{4} \zeta(2) + \frac{71}{192} \bigg)
                     + C_F C_A \bigg( \frac{1}{2} \zeta(2)
\label{mtosmtms}\\
&&
\hphantom{M_t 1+}
                     - \frac{1111}{384}
                     - \frac{1}{4} \pi^2  \log 2
                     + \frac{3}{8}\zeta(3) \bigg)
                     + C_F \bigg(
                     - \frac{3}{4} \zeta(2) + \frac{3}{8} \bigg)
\nonumber\\
&&
\hphantom{M_t 1+}
                     + C_F^2 \bigg(
                       \frac{1}{2} \pi^2 \log 2
                     - \frac{3}{4}\zeta(3)
                     - \frac{15}{8} \zeta(2)
                     + \frac{7}{128} \bigg) \Bigg)
         \Bigg] \nonumber
\eeqa
the result is easily expressed in terms of the OS mass:
\beqa
\lefteqn{\drho_{OS} = 3 X_t}                \label{drhoos}
\\ && \Bigg\{ 1
            + \frac{\as}{4\pi} C_F \bigg( -2 - 4 \zeta(2)\bigg)
\nonumber \\ & &
       + \left(\frac{\as}{4\pi}\right)^2\Bigg[
         C_F  \bigg( \frac{188}{3} + 486 S_2 - \frac{512}{3} \zeta(3)
                                  - \frac{100}{3} \zeta(2) \bigg)
\nonumber \\ & &
       + C_F^2  \bigg( -\frac{238}{9} - 1053 S_2
                + 4 D_3 + \frac{1012}{3} \zeta(3)
              + 4 \zeta(4)
\nonumber\\ & &
\phantom{C_F C_A}
             - \frac{770}{9} \zeta(2)
             + 8 B_4 + 96\zeta(2)\log 2 \bigg)
\nonumber \\ & &
       + C_F C_A   \bigg( -\frac{49}{6} + \frac{1053}{2} S_2 - 2 D_3
                   - \frac{416}{3} \zeta(3) + 30 \zeta(4)
                   - \frac{98}{3} \zeta(2) - 4 B_4
\nonumber\\ & &
\phantom{C_F C_A}
                   - 48\zeta(2)\log 2
                   - \frac{22}{3} \los - \frac{44}{3}\zeta(2) \los \bigg)
\nonumber \\ & &
       + C_F n_f   \bigg(  - \frac{2}{3} - \frac{16}{3} \zeta(3)
                      + \frac{52}{3} \zeta(2)
                      + \frac{4}{3} \los + \frac{8}{3}\zeta(2) \los \bigg)
\nonumber
\Bigg]
\Bigg\}.
\eeqa
Here $M_t$ is the pole mass and $X_t = G_F M_t^2/8\sqrt{2}\pi^2$.
The residual $\log\mu$ terms are cancelled by the $\mu$-dependence
of $\as$.

At this point two consistency checks should be mentioned which were performed
in order to test the correctness of our result.
The first one is a different method of calculating $\drho$ respectively
the polarisation functions for the $W$ and $Z$ boson. It relies on
the axial Ward identity which connects the axial part of
the polarisation tensor with the pseudoscalar polarisation function.
(The double triangle diagram was not considered in this
context.)
If the fermions in the loop have the masses $m_1$ and $m_2$ the following
identity holds:
\beq
 q_{\mu}q_{\nu}\Pi^{\mu\nu,a}(q)=(m_1+m_2)^2\Pi^p(q)
                   +(m_1+m_2)<0|\bar{q}_1 q_1+\bar{q}_2 q_2|0>.
\eeq
The second term of the r.h.s. is independent of $q$ and therefore not
relevant in this context. The l.h.s. in lowest order
is already ${\cal O}(q^2)$.
$\Pi^p(q)$ was evaluated up to ${\cal O}(q^2)$.
Because of the different tensor structure, different recurrence
relations have to be applied to compute the $\rho$ parameter.
$\Pi^p$ has to be expanded in $q$ up to order
$q^2$. Additional propagators are generated
and different parts of the FORM programs become relevant.
In addition it was checked that the $q$ independent part of $\Pi^p$
cancels against the second term such that the r.h.s. indeed is of
${\cal O}(q^2)$.
$\drho$ as given in eq. (\ref{drhoms}) was reproduced with this method.

The second check is also connected with an expansion in the external momentum.
The polarisation tensor of the vector
bosons was expanded up to order $q^2$. The external
momentum was routed through the graphs in two different ways.
Again the same result was obtained for every diagram although the
intermediate steps are very different.

Substituting $C_A=3,\,C_F=4/3$ and $\mu^2=\mbar_t^2$ or $\mu^2=M_t^2$
in eqs. (\ref{drhoms}) and (\ref{drhoos}) respectively
a fairly compact form for $\drho$ is obtained:
\beqa
\lefteqn{\drho_{\scriptsize\msbar} = 3 x_t}
\\ && \Bigg\{ 1
       + \frac{\as}{4\pi} \bigg( 8 - \frac{16}{3} \zeta(2) \bigg)
\nonumber\\ & &
       + \left(\frac{\as}{4\pi}\right)^2\Bigg[
         n_f \bigg(  - \frac{50}{3} - \frac{64}{9} \zeta(3)
               + \frac{112}{9} \zeta(2) \bigg)
\nonumber\\ & &
       +  \frac{26459}{81} + 882 S_2 - \frac{8}{9} D_3
       - \frac{5072}{27} \zeta(3) + \frac{1144}{9} \zeta(4)
       - \frac{25064}{81} \zeta(2) - \frac{16}{9} B_4
\nonumber
\Bigg]
\Bigg\}
\\
\lefteqn{\drho_{OS} = 3 X_t}
\\ && \Bigg\{ 1
       + \frac{\as}{4\pi} \bigg( -\frac{8}{3} - \frac{16}{3} \zeta(2) \bigg)
\nonumber\\ & &
       + \left(\frac{\as}{4\pi}\right)^2\Bigg[
         n_f \bigg(  - \frac{8}{9} - \frac{64}{9} \zeta(3)
               + \frac{208}{9} \zeta(2) \bigg)
       +  \frac{314}{81} + 882 S_2 - \frac{8}{9} D_3
\nonumber\\ & &
       - \frac{4928}{27} \zeta(3) + \frac{1144}{9} \zeta(4)
    - \frac{26504}{81} \zeta(2) - \frac{16}{9} B_4 -\frac{64}{3}\zeta(2)\log 2)
\nonumber
\Bigg]
\Bigg\}
\eeqa
To evaluate these results numerically the
values for $B_4$, $S_2$, $D_3$ and the $\zeta$ functions are inserted.
For convenience of the reader we display
separately four different contributions:
({\it i}\,) the contribution from the double triangle diagrams related to
the axial anomaly \cite{AnsDomLea93},
({\it ii}\,) the contributions with exactly one fermion loop
which together with the previous one would give the result in the
``quenched'' approximation,
({\it iii}\,) the contribution from light quark loops proportional to
$n_l\equiv n_f-1$, and finally
({\it iv}\,), the contribution with the light quarks replaced by top quarks.
\beqa
\drho_{\scriptsize\msbar} &=&
       3 x_t \Bigg( 1 - 0.19325\, \api
\nonumber\\ & &
\phantom{3 x_t  11}
         + (-4.2072 + 1.4151 - 0.29652 n_l + 0.30516) \,
                                      \left(\api\right)^2\Bigg)
                   \label{drhomsnl}                \\
\drho_{OS} &=&
       3 X_t \Bigg( 1 - 2.8599\, \api
\nonumber\\ & &
\phantom{3 x_t  11}
             + (-4.2072 - 19.416 + 1.7862 n_l + 0.098035) \,
      \label{drhoosnl}
                                \left(\api\right)^2 \Bigg).
\eeqa
The coefficient in front of $n_l$ in eq. (\ref{drhoosnl}) is in
reasonable agreement with the numerical result in
\cite{SmiVol94}.
Eqs. (\ref{drhomsnl}) and (\ref{drhoosnl}) in particular the separation
of the various contributions allow the use or test of a variety
of optimization schemes which is left to the reader.

For the final result and after setting $n_l=5$ we obtain
\beqa
\drho_{\scriptsize\msbar} &=&
       3 x_t \left( 1 - 0.19325\, \api - 3.9696\, \left(\api\right)^2\right)
                                            \\
\drho_{OS} &=&
       3 X_t \left( 1 - 2.8599\, \api - 14.594\, \left(\api\right)^2 \right).
                              \label{drhoosnum}
\eeqa
The coefficient in front of $\as^2$ in the $\msbar$ result differs
significantly from \cite{AvdFleMikTar94}: $-3.969\ldots$ versus
$+0.07111\ldots$.
The OS results differ by the same amount $-14.59\ldots$ versus $-10.55\ldots$.

In \cite{AvdFleMikTar94} it was mentioned that the contribution from
the double triangle diagram, associated with the axial anomaly
\cite{AnsDomLea93}, alone
amounts to about $40\%$ of
the total three-loop correction.
Here a fraction of approximately
$30\%$ is still traceable back to this single diagram in the OS scheme.
In the $\msbar$ scheme the corrections are completely dominated
by this diagram.

Finally the numerical effect on the prediction for
$M_W$ and $\sineff$ from $\al, G_F$ and $M_Z$ will be discussed.
If sub-leading terms are neglected the following relations hold
\cite{ConHolJeg89}:
\beqa
M_W^2 & = & \frac{\rho M_Z^2}{2}\left(
            1+\sqrt{1-\frac{4A^2}{\rho M_Z^2}\frac{1}{1-\Delta\al}}
                                \right) \nonumber \\
\sineff & = & 1-\frac{M_W^2}{\rho M_Z^2} =
            \frac{1}{2}\left(
            1-\sqrt{1-\frac{4A^2}{\rho M_Z^2}\frac{1}{1-\Delta\al}}
                         \right). \nonumber
\eeqa
Here $A=\sqrt{\pi\al/\sqrt{2}G_F}=37.2802$ GeV, $\Delta\al\approx 0.06$
and $\rho = 1/(1-\drho)$.
The relative size of the one-, two- and
three-loop
corrections is given in Table \ref{table1}.
The first row indicates the size of $\drho$ itself.
Rows two and three give the relative contribution of the one-, two- and
three-loop corrections with respect to the Born result in the OS and $\msbar$
scheme.
The numbers are obtained with the following input data:
$\as(M_t^2)=0.1092$, $\mbar_t(M_t^2)=164$ GeV, $M_t=174$ GeV,
$M_Z=91.188$ GeV and
$G_F=1.16639\,10^{-5}$ GeV$^{-2}$.
The value of $\as(M_t^2)=\as^{(6)}(M_t^2)$ was obtained from
$\as^{(5)}(M_Z^2)=0.120$ with the help of the three-loop $\beta$ function
\cite{TarVlaZha80} and the matching condition between $\as^{(5)}(\mu^2)$
and $\as^{(6)}(\mu^2)$ at the scale $\mu^2=M_t^2$
\cite{LarRitVer94}.
(The numbers in brackets
indicate thereby the numbers of active flavours.)
The $\msbar$ top mass $\mbar_t(M_t^2)$ was derived from
eq. (\ref{mtosmtms}). This value serves as starting point in order to
calculate $\mbar_t(\mu^2)$ using the corresponding three-loop
evolution equation
\cite{Tar92}.
(Unlike the coupling constant the running top quark mass is only defined
in full $n_f=6$ theory.)

\begin{table}[h]
\renewcommand{\arraystretch}{1.3}
\begin{center}
\begin{tabular}{c}
\begin{tabular}{|l||r|r|r|}
\hline
OS        & +1-loop & +2-loop & +3-loop \\
\hline
\hline
$\drho$                 & 0.00949  &  0.00854   &  0.00838 \\
$\delta M_W/M_W$        & 0.00682  &  0.00614   &  0.00601        \\
$\delta\sineff/\sineff$ & -0.01349 &  -0.01216  &  -0.01192       \\
\hline
\end{tabular}
\\
\\
\begin{tabular}{|l||r|r|r|}
\hline
$\msbar$  & +1-loop & +2-loop & +3-loop \\
\hline
\hline
$\drho$                 & 0.00843  & 0.00837   & 0.00833 \\
$\delta M_W/M_W$        & 0.00605  & 0.00601   & 0.00598  \\
$\delta\sineff/\sineff$ & -0.01200 & -0.01191  & -0.01186  \\
\hline
\end{tabular}
\end{tabular}
\end{center}
\caption{\label{table1} Numerical results including successively
higher orders.}
\end{table}

It is interesting to
mention that the scheme dependence decreases enormously when taking
higher loop corrections successively into account.

One observes that the three-loop corrections are (at least for $M_W$)
not at all small compared with the two-loop QCD contribution.
Furthermore  they are approximately of the same order of magnitude
as the two-loop electroweak result because for $m_H/m_t=1.5$ the contribution
from the two-loop electroweak correction amounts $-0.018\%$
\cite{AvdFleMikTar94,Bar92FleTarJeg93}
to be compared with the
three-loop QCD corrections of $-0.013\%$.
It is also possible to translate the ${\cal O}(\as^2)$ corrections
directly in a change of the top mass contained in $X_t$.
In the on-shell scheme this corresponds to a change
of approximately $-1.5$ GeV.

\begin{figure}[b]
 \begin{center}
 \begin{tabular}{c}
   \epsfxsize=12.0cm
   \leavevmode
   \epsffile[55 190 520 670]{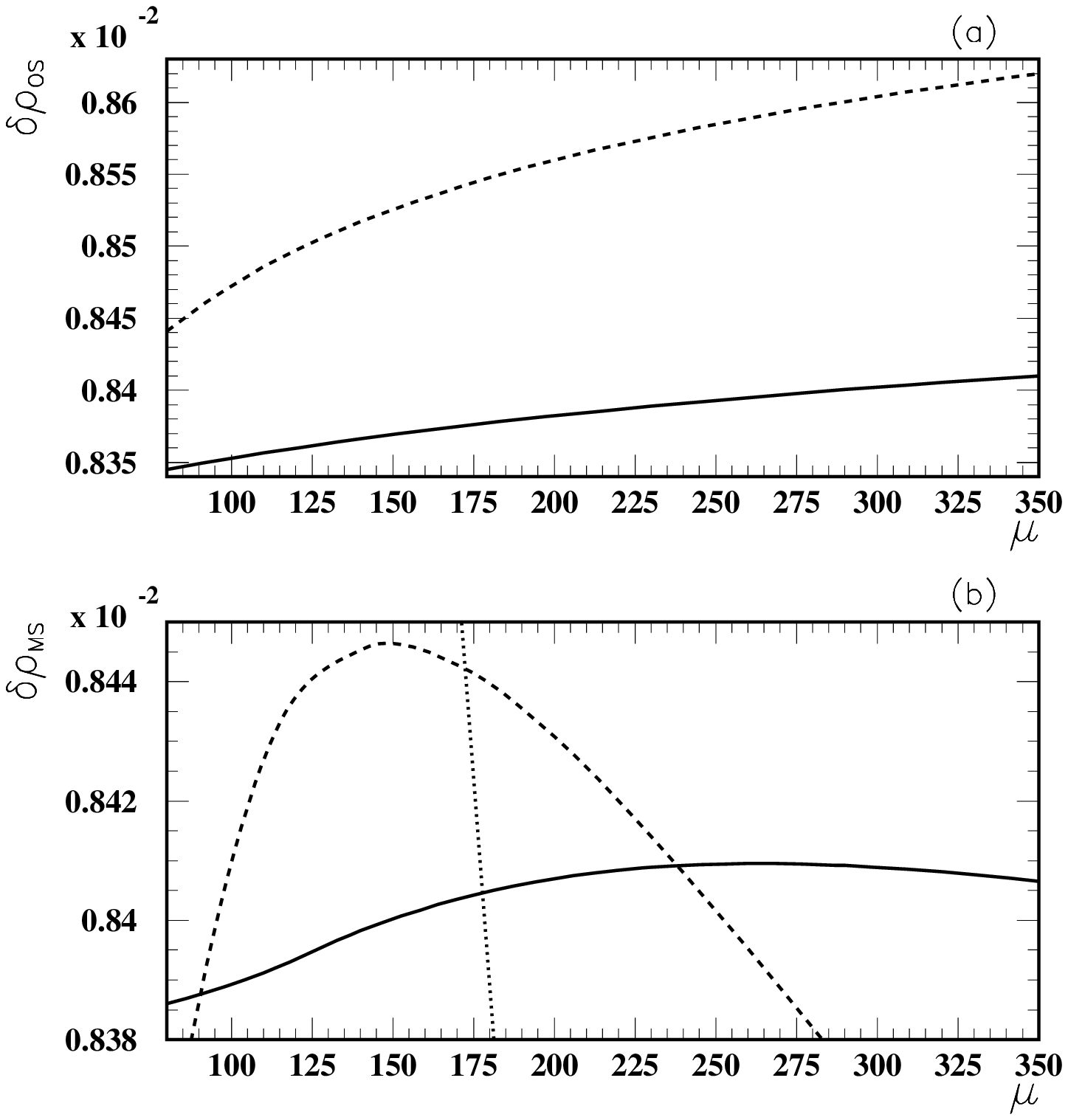}
 \end{tabular}
 \caption{\label{fig} Renormalization scale dependence of
        $\drho_{OS}$ (a) and $\drho_{\scriptsize\msbar}$ (b).}
 \end{center}
\end{figure}

The dependence of the result on the renormalization scale $\mu^2$
is shown in Figs. \ref{fig}a and \ref{fig}b
for $\drho_{OS}$ and $\drho_{\scriptsize\msbar}$
respectively. The same input parameters have been used as before.
The dotted line in Fig. \ref{fig}b gives the one-loop prediction
(which is constant for $\drho_{OS}$ and completely off-scale in
Fig. \ref{fig}a), dashed
and solid lines represent the two- and three-loop results.
The prediction is clearly stabilized through inclusion
of higher orders.

Another possibility would be to absorb the $\as^2$ contribution in the choice
of an effective scale of the $\as$ correction:
\beq
\drho_{OS} =
       3 X_t \left( 1 - 2.8599\, \frac{\as((0.302M_t)^2)}{\pi}\right).
\eeq

To summarize: The evaluation of the three-loop QCD correction
to the $\rho$ parameter has been repeated with a result different from the
one of \cite{AvdFleMikTar94}.
The numerical difference is sizeable.

\vspace{5ex}
{\bf Acknowledgments}

\noindent
We would like to thank G. Passarino and A. Sirlin for discussions and
A. Czarnecki for advice in the numerical evaluation of Feynman integrals.
One of the authors (M.S.) would like to thank B.A. Kniehl for the
occasion to present the results of this paper at the Ringberg
Workshop on ``Perspectives for electroweak interaction in $e^+e^-$
collisions'', February 5-8, 1995.

%
%

\end{document}